\documentclass{aa} 

\usepackage{natbib}
\bibpunct{(}{)}{;}{a}{}{,} 

\usepackage{graphicx}

\newcommand{\FeIX}{\ion{Fe}{ix}}
\newcommand{\FeXII}{\ion{Fe}{xii}}

\newcommand{\kms}{km~s$^{-1}$}
\newcommand{\degree}{\ensuremath{^\circ}}

\begin{document}

\authorrunning{Panesar, Innes, Tiwari, Low}
\titlerunning{Tornado triggered by Flares}
\title{A solar tornado triggered by flares?}

\author {N.~K. Panesar\inst{1,}\inst{2}
\and D.~E. Innes\inst{1}
\and S.~ K. Tiwari\inst{1}
\and B. C. Low\inst{3}}

\institute{Max-Planck Institut f\"{u}r Sonnensystemforschung,
 Max-Planck-Str. 2, 37191 Katlenburg-Lindau
\and Institut f\"{u}r Astrophysik, Georg-August-Universit\"{a}t G\"{o}ttingen, Friedrich-Hund-Platz 1, D-37077 G\"{o}ttingen
\and High Altitude Observatory, National Center for Atmospheric Research, P.O. Box 3000, Boulder, CO 80307, USA}

\offprints{N.~K. Panesar \email{panesar@mps.mpg.de}}



  \abstract
   {Solar tornados are dynamical, conspicuously helical magnetic structures mainly observed as a prominence activity.}
  {We investigate and propose a triggering mechanism for the solar tornado observed in a prominence cavity by SDO/AIA on September 25, 2011.}
   {High-cadence EUV images from the SDO/AIA and the Ahead spacecraft of STEREO/EUVI are used to correlate three flares in the neighbouring active-region (NOAA 11303), and their EUV waves, with the dynamical developments of the tornado. The timings of the flares and EUV waves observed on-disk in 195\AA\ are analyzed in relation to the tornado activities observed at the limb in 171\AA.}
   { Each of the three flares and its related EUV wave occurred within 10 hours of the onset of the tornado. They have an observed causal relationship with the commencement of activity in the prominence where the tornado develops. Tornado-like rotations along the side of the prominence start after the second flare. The prominence cavity expands with acceleration of tornado motion after the third flare.}
{Flares in the neighbouring active region may have affected the cavity prominence system and triggered the solar tornado.
A plausible mechanism is that the active-region coronal field contracted by the
`Hudson effect' due to the loss of magnetic energy as flares. Subsequently
the cavity expanded by its magnetic pressure to fill the surrounding low
corona. We suggest that the tornado is the dynamical response
of the helical prominence field to the cavity expansion.}
   \keywords{Sun: chromosphere -- Sun: prominences -- Sun:
   flares -- Sun: coronal cavity }

\maketitle

\section{Introduction}

Prominences consist of relatively large, cool and over dense plasma seen in
the lower corona above the solar limb \citep{Martin98,tand95,Mackay10}. Their
structure and composition are exceedingly complicated. The plasma mainly
resides in highly tangled magnetic fields \citep{Balle10}. Coronal cavities are
often observed to have cooler prominence plasma at their bases \citep{Hudson99, Gibson06,
Regnier11}. The cavity is a region of relatively low density, high
temperature plasma \citep{gib10,hab10}. \cite{ber11} have proposed that the prominence and its cavity are a
form of magneto-thermal convective structure, macroscopically stable but internally in a constant state of
ubiquitous motions \citep{low12a,low12b}.

\begin{figure*}
   \sidecaption
  \includegraphics[width=14.9cm]{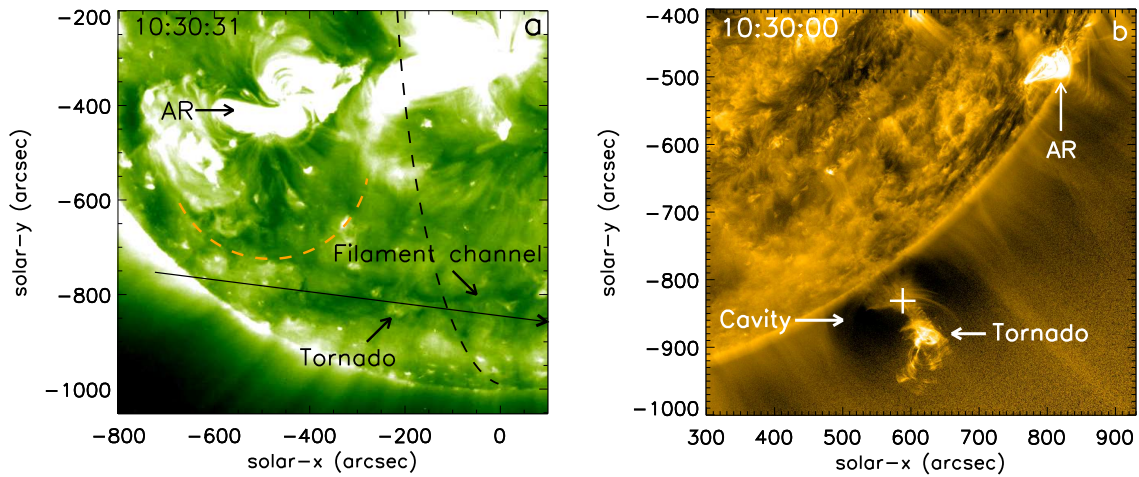}
     \caption{The region around a tornado on 25 September 2011: a) EUVI-A 195\AA; b) AIA 171\AA\ images.
  On the EUVI image, the black dashed line shows the position of the AIA limb,
  and the long diagonal arrow is the epipolar line for the prominence position marked with
'+' on the AIA image. The orange dashed line outlines the position of the southern edge of the AR corona.} \label{over}
\end{figure*}

  Viewed at the limb, quiescent prominences mostly appear as curtains of vertical
thread-like structures \citep{ber08}. Occasionally they look like tornados with rotations
along, or of, their magnetic structures \citep{pet43,liggett84}. At first
glance it is not obvious why tornados do not erupt but continue rotating for
several hours before quietening down. Recently, images from Atmospheric
Imaging Assembly (AIA) on the Solar Dynamic Observatory (SDO), have allowed
the study of solar tornados \citep{li12,yang12,jia12}. The driving mechanism may be
a coupling and expansion of a twisted flux rope into the coronal cavity
\citep{li12} and/or related to photospheric vortices at the footpoint of the
tornado \citep{attie09,sven12,yang12}.

This letter is a study of the impressive solar tornado observed by AIA on 25 September
2011. \citet{li12} described the formation and disappearance of the related
prominence by analysing SDO data from 24-26 September. The evolution of the
tornado was attributed to the expansion of helical structures into the
cavity. But the reason behind the expansion remained an open question.

Here we analyze observations on the mechanism leading to the
expansion of the prominence. We show using Solar TErrestrial RElations Observatory (STEREO)
Extreme UltraViolet Imager (EUVI) observations that three strong flares in a
neighbouring active region coincided with phases of the tornado activation.
Each flare was associated with a Coronal Mass Ejection (CME) and EUV wave.
Particularly, after the third flare, there was slow but significant expansion of the
overlying prominence cavity together with more rapid rotation at the top of
the prominence. We describe the observations in Section 2. The evolution of
the tornado, its relationship to the solar flares and EUV waves are described
in Section 3. In Section 4, we summarize our observations and speculate on
the link between the flares and the cavity expansion.

\section{Observations}
The solar tornado on 25 September 2011 was observed on the south west limb by
SDO/AIA \citep{Lemen11}. In images from EUVI on the Ahead spacecraft of
STEREO \citep{Howard08}, it appeared on the solar disk around 15\degree\ E.
The separation angle between SDO and STEREO-A was 103\degree. We study the
tornado by combining observations from the two directions.

AIA takes high spatial resolution (0.6\arcsec\ pixel$^{-1}$) full disk images
with a cadence of 12~s. For the analysis, we select images from the 171\AA\
channel which is centered on the \FeIX\ line formed around 
0.63 MK. Images from this channel show both hot, bright emitting
and cold, dense absorbing parts of the prominence, as well as the dark,
low density cavity and the surrounding coronal loops. To enhance the
visibility of the cavity and faint loops, we removed an average background
from the 171\AA\ images by taking the median of two months data - September
and October, 2011.

During the period studied, 01:30-13:30~UT, SDO was in eclipse from
06:02-07:13~UT. This gap is covered by the SWAP instrument on PROBA-2
\citep{halain10}. SWAP provides a full Sun 174\,\AA\ image every 2-3~min with a
spatial resolution of 3.16\arcsec\ pixel$^{-1}$. The images are not as
detailed as AIA images but are essential for checking the behaviour of the
prominence during the data gap.

The EUVI-A 195\AA\ images have a time cadence of 5~min and resolution of
1.6\arcsec\ pixel$^{-1}$. The 195\AA\ emission is mainly \FeXII\ formed at $\sim$1.2~MK. To confirm the prominence position in the EUVI images, we
obtained the three-dimensional coordinates of the prominence with the routine
SCC$\_$MEASURE \citep{thom09} available in the SolarSoft library.

\subsection{Overview}

    In Fig.~\ref{over} we show the EUVI-A 195\AA\ and AIA 171\AA\ images of the
    filament/prominence and the active region NOAA 11303 (marked with an arrow)
    as seen from the two angles. The position of the cross on the stem of the prominence lies along the
    epipolar line drawn as a long black arrow on the EUVI image.
    The dark lane crossed by the epipolar line is the filament channel
    and the tornado is the bright region that coincides with the epipolar line
    and the filament channel, indicated with a black arrow. The core
    of the active region is separated by 300\arcsec\ from the filament
    channel. However, the southern edge of the active region corona (outlined by orange dashed line) reaches to within 50\arcsec\ of the filament
    channel.

\subsection{The three flares}

\begin{figure*}
   \centering
   \includegraphics[width=0.9\linewidth]{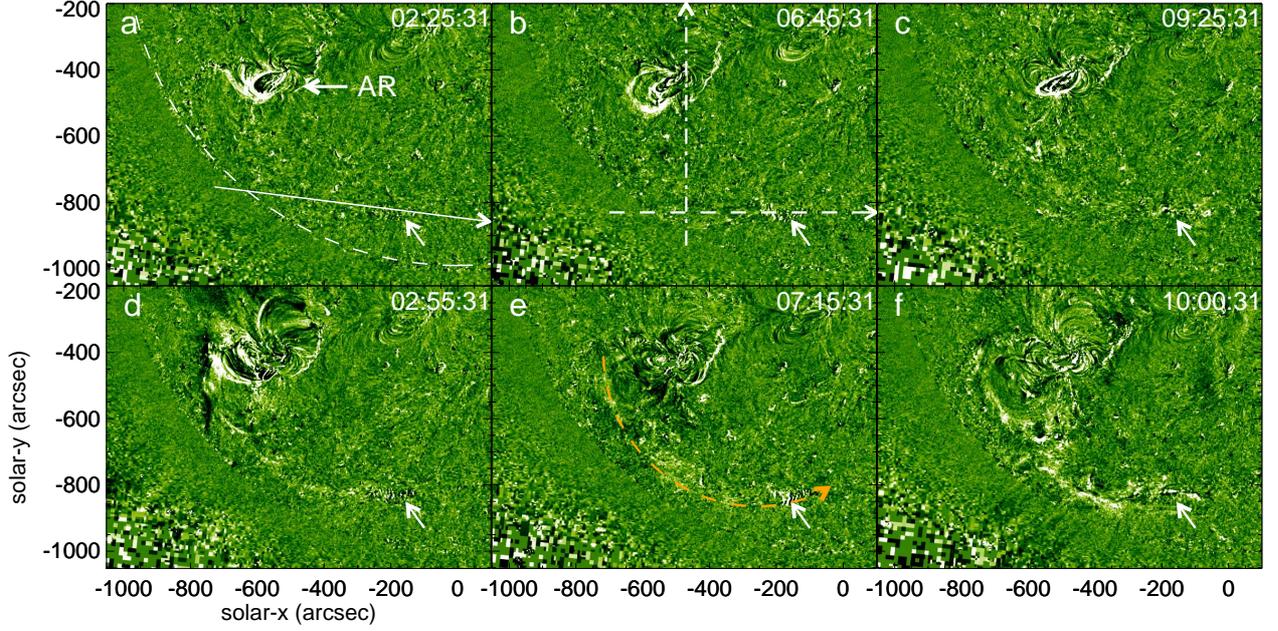}
      \caption{ EUVI-A 195\AA\ running ratio images showing the onset of the three flares (a,b,c) and their associated EUV waves (d,e,f).
      In (a) the epipolar line shown in Fig.~\ref{over}a is drawn as a long white arrow, and the solar limb is drawn
as a dashed white line. The small white arrow in all images points to the
tornado site. The dot-dashed/dashed lines in (b)
 mark the positions of time series shown in Fig.~\ref{stcut},
 and the orange dashed line along the EUV wave front in (e) the position of the time series in Fig.~\ref{wavecut}.
      }\label{waves}
\end{figure*}

There were flares at approximately 02:45, 07:00, 09:40~UT from the
nearby active region. They were all associated with CMEs and EUV waves. The
third was GOES class M1.4 according to the NOAA flare
catalogue\footnote{ftp://ftp.ngdc.noaa.gov/STP/space-weather/solar-data/solar-features/solar-flares/x-rays/goes}.
The GOES class of the first two are not given in the catalogue. We therefore
checked the hard X-ray quicklook images from the Reuven Ramaty High Energy
Solar Spectroscopic Imager (RHESSI). Unfortunately they both had their peaks
during RHESSI data gaps. The first RHESSI image in the flares' decay phase
showed that these flares were the brightest RHESSI sources. At the
time of the first and second flares the GOES 1-8\AA\ fluxes reached the M4.4
and M1.0 level respectively, so it is possible that these too were
M-class flares.

The flares' onsets and related EUV waves can be seen in the 195\AA\
running ratio images shown in Fig.~\ref{waves}. The running ratio images are
the log of the intensity at the time shown on the images divided by the
intensity 5~min earlier. Because this reflects relative changes the faint EUV
waves show up against bright inactive active regions. The faint EUV
wavefronts sweep over the filament channel about half an hour after the flare
(bottom row). The small white arrows point to the tornado site where there
was increased activity along the filament channel after the flares.

\begin{figure}
 \centering
\includegraphics[width=\linewidth]{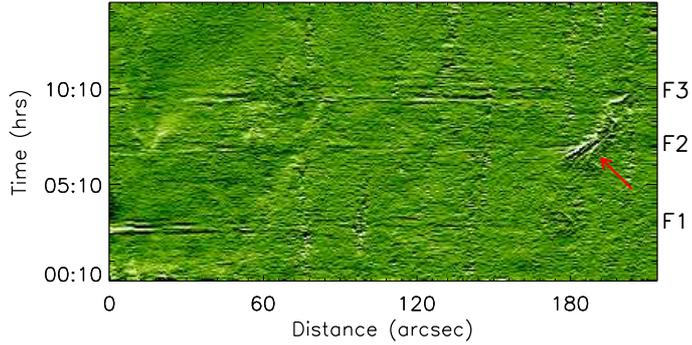}
\caption{ EUVI 195\AA\ running ratio space time image along the dashed line
in Fig.~\ref{waves}e. Three flares are labelled as F1, F2 and F3. The red
arrow points to the filament. }\label{wavecut}
\end{figure}

When we take a time series along the orange dashed line drawn on
Fig.~\ref{waves}e, we are able to see the relation between the EUV waves and
the filament activity (Fig.~\ref{wavecut}). In this time series image,
   the waves associated with the three flares are marked by F1, F2 and F3 and a red arrow points to the filament activity.
   We note that the filament activity associated with F2 increased before
   the arrival of the EUV wave. We also notice that each EUV wave front was followed by a series of oscillations
    with a period of 20 min at the edge of the active region and along the filament
   channel. Similar post-EUV cavity oscillations, seen at the limb in AIA images,
   have been reported by \citet{liu12}. EUVI images have a much lower cadence than AIA, and may not be resolving the wave trains.

\begin{figure}
 \centering
 \includegraphics[width=\linewidth]{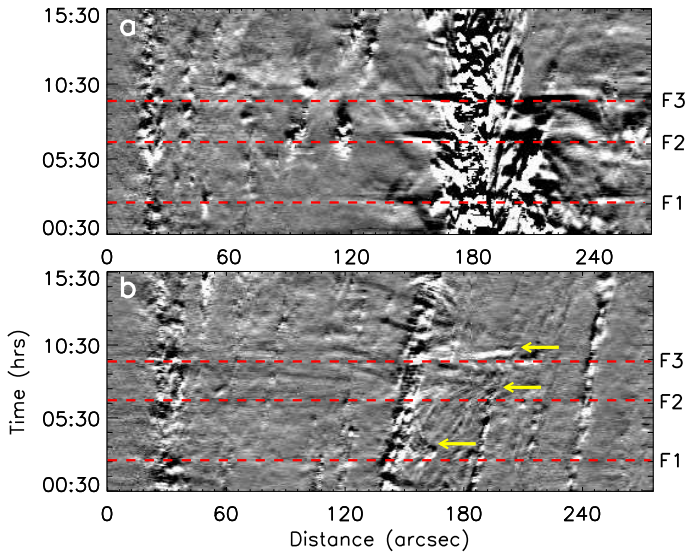}
 \caption{
 EUVI 195\AA\ running difference time series along (a) the vertical and (b) the horizontal line
 in Fig.~\ref{waves}b. The red dashed lines are drawn at the flare F1, F2, and F3 times. In b) the yellow arrows point to the
 changes in filament structure after the flares.
 }\label{stcut}
\end{figure}

 As mentioned above, the EUV wave reached the filament channel after the filament activation.
 To investigate whether the activation could have been triggered by the flare,
 we took time series through the flare sites (vertical line Fig.~\ref{waves}b)
 and along the filament channel across the activated filament (horizontal line in Fig.~\ref{waves}b).
  We observe that there are perturbations in the filament plasma immediately after each flare (yellow
arrows in Fig.~\ref{stcut}b). Large-scale changes along the filament channel started after F2 and
increased further after the F3.

The reaction of the prominence to the flares is illustrated in
Fig.~\ref{sdotimeseries}. The positions of the three time series are drawn on
Fig.~\ref{sdocut}, an AIA image of the active region and tornado. The
oscillations of the prominence at the time of F1 (Fig.~\ref{sdotimeseries}a) are consistent with a flare
trigger from the north since it moved first towards the pole.
The wobbling in the prominence after the flare is also clearly visible in the
movie6a (Fig.~\ref{sdocut}).

\begin{figure}
   \centering
   \includegraphics[width=0.9\linewidth]{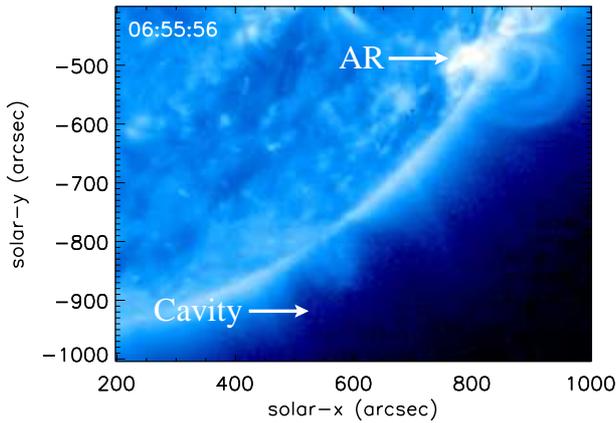}
   \caption{SWAP 174\AA\ intensity image from 25 September 2011.
   The arrows show the positions of the cavity
   and the active region (AR). This frame is taken from the movie-"MOVIE5.mp4".
   }
\label{sp}
\end{figure}

\begin{figure}
   \centering
   \includegraphics[width=0.9\linewidth]{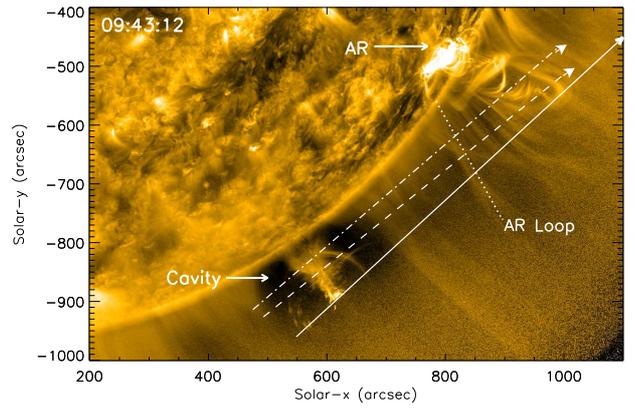}
   \caption{AIA 171\AA\ intensity image.
   The diagonal long arrows show the positions of the time series images through the prominence/tornado, cavity,
   active region hot loop (AR loop) and active region (AR) shown in Fig.~\ref{sdotimeseries}. This is the frame from the movie-"MOVIE6b.mp4".
   "MOVIE6a.mp4" shows the evolution of this region at earlier times.
   }
\label{sdocut}
\end{figure}

\begin{figure}
   \centering
   \includegraphics[width=\linewidth]{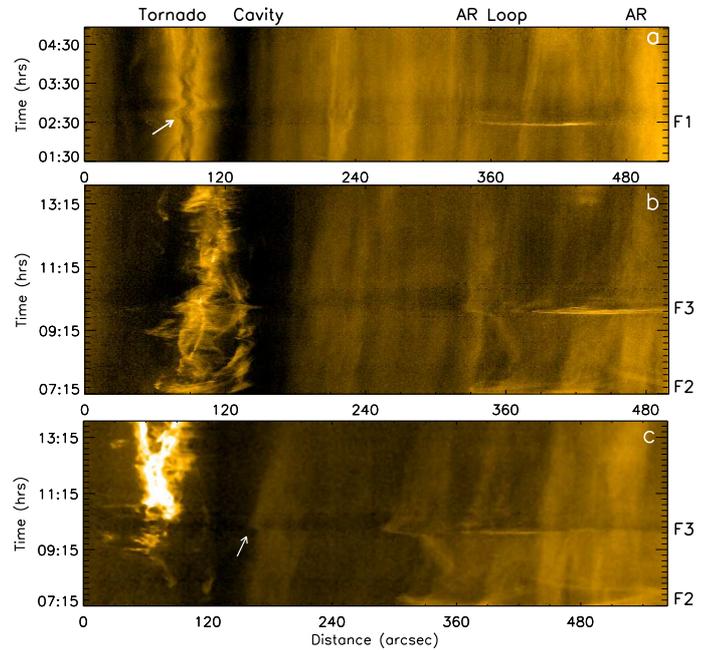}
   \caption{AIA 171\AA\ intensity time series along the diagonal arrows in
   Fig.~\ref{sdocut}: a) oscillations of the prominence stem triggered by F1 taken along bottom white arrow;
   b) activity triggered by F2 and F3 taken along the middle white arrow;
   c) tornado activity and cavity changes triggered by F2 and F3 taken along top arrow.
    The arrow at the time of the flare ($\sim$09.40 UT) points to the movement of
   the cavity boundary.}\label{sdotimeseries}
\end{figure}

The second flare, F2, was at $\sim$06:45~UT when SDO was in eclipse. Since this
phase is important for understanding the tornado, we checked
 the SWAP data to see if the prominence activation started before F2.
The SWAP data are shown in the movie attached with Fig.~\ref{sp}.
 Although the SWAP images are not as
sharp as AIA images, they show quite clearly that
 increased prominence activity occurred after F2 and that
the first AIA images after eclipse caught the expansion of the flare loops
and the rapid growth in prominence activity.
  The middle image in Fig.~\ref{sdotimeseries}b, taken along the dashed line in
Fig.~\ref{sdocut}, shows an arm of prominence plasma reaching out in the
direction of the active region. There is also a very slight expansion of the
cavity boundary towards the active region.

The biggest change in the cavity occurred after F3 when the bright and
complex top part of the prominence started to rotate faster.
The movie6b attached to Fig.~\ref{sdocut}, shows the evolution of the tornado and
flare from F2 to a couple of hours after F3.
  The time series
in Fig.~\ref{sdotimeseries}c, taken along the top white arrow in
Fig.~\ref{sdocut}, shows the main features. The flare erupted at 09:20~UT. The
 active region loops started expanding at 09:40:48~UT.
 At 09:50:13~UT there was a sharp contraction, approximately 10~Mm, of
the observed cavity. The cavity started expanding back towards the active region after
$\sim$10:00~UT. It grows from 140~Mm to 167~Mm in 3.5 hours. At the same
time, the prominence grew in height and the bright and complex head of the
prominence started to rotate faster, forming the tornado. The
rotations at the top of the prominence/tornado lasted $\sim$3.5 hours with a
speed from 55~\kms\ to 95~\kms\ \citep{li12}.

\section{Discussion}

A solar tornado was observed by AIA and EUVI-A on 25 September 2011. The
tornado was reported by \cite{li12}, who attributed it to the growth of a
helical prominence system. The reason for the prominence growth remained
unanswered. After careful observation of the event using AIA full disk
images, we noticed that AR 11303, to the north of the prominence cavity
was flaring and suspected it of having influenced the prominence in
some way. We observed three flares, and they were all associated with EUV
waves that swept over the prominence cavity. Each time the cavity was
buffeted by the wave, the prominence plasma became more active. After the
third flare a tornado had developed at the top of the prominence.

After the first flare, there were clear oscillations along the prominence
stem. The second flare caused plasma to move along arm-like extensions
projecting out from the stem in the direction of the active region. These extensions appear to subsequently rotate
 back towards the prominence before the third flare. After the
third flare the tornado became very strong and was stable for about
3.5~hours. The third flare seemed to have the biggest impact on the
surrounding plasma or at least produced the most visible (at 195\AA) EUV
wave. It also caused large active region loops to move towards the edge of the
prominence cavity. Afterwards the loops swayed back and the cavity started to
slowly ($\sim$ 2.5~\kms) but visibly expand over the next 3.5~hours. During this time
the prominence rose into the cavity creating the tornado.

To understand the relationship between the flare, the prominence, the cavity
and the tornado there are a number of effects that need to be considered. The expansion
of the cavity is clearly associated with the rise of the prominence plasma.
One possible explanation for the expansion is that the loss of free magnetic energy from the active region by flares and
CMEs \citep{zha05} resulted in a contraction of the active region field,
the Hudson effect \citep{hud00,zha03,jan07}. Subsequently the surrounding fields, including the cavity field, expanded to fill the vacated space.

\begin{figure}
   \sidecaption
   \includegraphics[width=5.535cm]{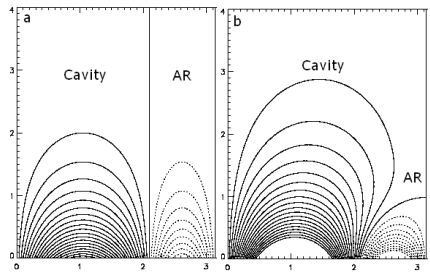}
   \caption{Idealized 2D Cartesian magnetic fields of the cavity and active region (a) before
   and (b) after the flare, in a domain $0<y<\pi, 0<z<\infty$ with $z$ denoting coronal height.}\label{f1}
\end{figure}

An idealized illustration of the Hudson effect, in terms of the force-free equilibrium between two bipolar fields
before and after one of the fields has lost its free energy, is given in Fig.~\ref{f1}. Initially the active region field on the
right has excess free energy, and the cavity field on the left is taken to be potential in order to simplify the calculation.
During the eruption, the active region field
loses free energy, decreasing its pressure since the magnetic energy density $B^2$ of a field ${\bf B}$ is also the
magnetic pressure. This leads to an expansion of the cavity field to restore pressure balance in the system.

We consider the 2D Cartesian domain $0<y<\pi$, $0<z<\infty$, with $x$ as an
ignorable coordinate, taking the boundary to be a rigid perfect electrical conductor for simplicity.
The initial state in Fig.~\ref{f1}a is a continuous global solution for the magnetic field, ${\bf B}={\bf B}_{initial}$ of the force-free equations
\begin{equation}
\label{fff}
\nabla \times {\bf B}=\alpha{\bf B}=0; ~~~~ \nabla \cdot {\bf B} = 0 ,
\end{equation}
\noindent
describing a bipolar ($\alpha=0$) potential field representing the cavity that occupies the partial domain $0<y<2\pi/3, 0<z<\infty$. This field continues into a constant-$\alpha$
force-free field representing the active region that occupies the complementary partial domain $2\pi/3<y<\pi, 0<z<\infty$. Their interface $y=2\pi/3$,
is in force balance because the fields on its two sides exert equal magnetic pressure. Suppose the electric current of the
constant-$\alpha$ force-free field is removed from the partial domain
 $2\pi/3<y<\pi, 0<z<\infty$ to represent a flare-like loss of free energy.
The distribution of the normal component of the field along the rigid boundary and the magnetic foot-points on the base $z=0$ cannot
change. This boundary condition then determine the unique end-state, ${\bf B}_{end}$, with $\alpha=0$ everywhere in the domain (Fig.~\ref{f1}b).
The former bipolar field on the right has contracted downward
and withdrawn into a partial domain of a finite height while the other bipolar field now occupies the whole infinite space above that height.
The fluid-interface between the two fields is now an arc of finite length. These mathematical solutions, given in the Appendix, serve
to conceptually make specific a basic effect in the complex processes of a real energy release.

In addition to the active region contraction, the CME/flare eruptions generated hydromagnetic disturbances,
seen as EUV waves, that perturbed the surrounding low corona. They evidently caused swaying in the prominence and
oscillations along the cavity boundary. These perturbations may have been intense enough to partially destabilize the cavity-prominence system,
thereby contributing to or possibly even initializing the tornado-like activity.

Our observational study supports the interpretation that the CME/flare activities in the active region
had a causal relationship with the nearby tornado. It is clear that the tornado was dynamically
associated with the expansion of the prominence's helical field and its cavity. It is an open question
as to whether the Hudson effect drove the cavity expansion and the helical field was responding
dynamically to the expansion, or, the helical field loses its meta-stability as the result of hydromagnetic
disturbances from the CME/flares from the active region. This physical question is worthy of MHD
time-dependent simulation. Evidence of the Hudson effect has been reported in recent solar observational studies
\citep[e.g.][]{sun12,riu09}. Our study shows the importance of this effect in relation to CMEs, flares and prominences,
and provides observational motivation for theoretical MHD modeling to address the specific questions posed by this relationship.

\begin{acknowledgements}
We are obliged to the SDO/AIA, STEREO/EUVI and SWAP teams. NKP
acknowledges the facilities provided by the MPS. We are thankful to Joan Burkepile, Thomas Berger
and Wei Liu for their useful inputs. The National Center for Atmospheric Research is sponsored by the US National Science Foundation.
\end{acknowledgements}

%
%
\appendix

\onecolumn
\section{Mathematical Illustration of the Hudson Effect}

We give the mathematical solutions for 2D magnetic fields in Fig.~\ref{f1}, which are straight forward to construct by standard techniques. Use the representation
\begin{equation}
\label{B}
{\bf B} = \left[ Q, {\partial A \over \partial z}, -{\partial A \over \partial y} \right] ,
\end{equation}
\noindent
in terms of the scalar flux-function $A(y, z)$ and the component $B_x=Q(y, z)$ corresponding to the ignorable coordinate $x$. The initial field ${\bf B}_{initial}$ in Fig. 7a is given by $A=A_{initial}$ and $Q=Q_{initial}$ where
\begin{eqnarray}
A_{initial}&=&A_{pot} = 2 \sin \left({3 \over 2} y \right) \exp(-{3 \over 2} z); ~~ Q_{initial} \equiv 0 , ~~ \mathrm{in} ~~ 0<y<2\pi/3, 0<z<\infty \nonumber \\
A_{initial}&=& A_{fff}=-\sin \left(3 y \right) \exp(-{3 \over 2} z) ; ~~ Q_{initial} = \sqrt{27 \over 3} A, ~~ \mathrm{in} ~~ 2\pi/3<y<\pi, 0<z<\infty .
\end{eqnarray}
\noindent
The end-state field ${\bf B}_{end}$ is the everywhere-potential field with $A=A_{end}$ where
\begin{eqnarray}
A_{end}&=&\sum_{n=1}^{\infty} a_n \sin ny \exp(-nz) , \\
a_3 &=& {1 \over 3} ~~\mathrm{and} ~~ a_n = {2 \times 3^4 \over \pi} \sin {2 \pi n \over 3} {1 \over (4 n^2 - 9) (n^2 - 9) } , \mathrm{~~for ~ n \ne 3} ,
\end{eqnarray}
\noindent
with $Q \equiv 0$. Both ${\bf B}_{initial}$ and ${\bf B}_{end}$ have the same distribution of normal field component along the boundary of the domain $0<y<\pi$, $0<z<\infty$.

\end{document}